\documentclass[aps,prl,twocolumn,superscriptaddress,longbibliography]{revtex4-1}

\usepackage[colorlinks=true,citecolor=blue]{hyperref}
\usepackage{graphicx}
\usepackage{amsmath}
\usepackage{amssymb}

\DeclareGraphicsExtensions{.png,.jpg,.eps}
\usepackage{xcolor}

\def\br{{\bold{r}}}

\def\m1{{^{-1}}}

\renewcommand{\H} {\mathcal{H}}
\newcommand{\dis} {d} 
\begin{document}

\author{Aline Ramires}
\affiliation{Condensed Matter
Theory Group, Paul Scherrer Institute, CH-5232 Villigen PSI, Switzerland}

\author{Jose L. Lado}
\affiliation{Department of Applied Physics, Aalto University, 00076 Aalto, Espoo, Finland}

\title{
Emulating  heavy fermions in twisted trilayer graphene 
}

\begin{abstract}
	Twisted van der Waals materials have been shown to host a variety of tunable electronic structures. Here we put forward twisted trilayer graphene (TTG) as a platform to emulate heavy fermion physics. We demonstrate that TTG hosts extended and localized modes with an electronic structure that can be controlled by interlayer bias. In the presence of interactions, the existence of localized modes leads to the development of local moments, which are Kondo coupled to coexisting extended states. By electrically controlling the effective exchange between local moments, the system can be driven from a magnetic into a heavy fermion regime, passing through a quantum critical point, allowing to electrically explore a generalized Doniach phase diagram. Our results put forward twisted graphene multilayers as a platform for the realization of strongly correlated heavy fermion physics in a purely carbon-based platform.
\end{abstract}

\date{\today}

\maketitle

Twisted van der Waals multilayers have recently risen as a new family of systems allowing for the realization of multiple exotic quantum states of matter \cite{Li2009,correlatedCao2018,superCao2018,Lu2019,Yankowitz2019,Rickhaus2018,Cao2020,2020arXiv200505373R,Serlin2019,Liu2020}.
Their versatility stems from the control of the strength
of electronic correlations and emergent
topological properties by means of the twist angle between
the different layers \cite{PhysRevB.82.121407,PhysRevB.84.195421,PhysRevLett.99.256802,Bistritzer2011,PhysRevB.92.075402,PhysRevLett.119.107201,PhysRevLett.120.266402,PhysRevLett.121.146801,2021arXiv210306313D}.
As paradigmatic examples,
twisted graphene bilayers have realized
unconventional superconductivity \cite{superCao2018}, topological networks \cite{Rickhaus2018},
strange metals \cite{PhysRevLett.124.076801}, and Chern insulators \cite{Serlin2019}.
More complex twisted graphene multilayers, such as twisted bi-bilayers \cite{Cao2020,Liu2020} and trilayers \cite{Chen2020tri,2020arXiv201201434P,2020arXiv201202773H}, have 
provided additional platforms to realize similar physics as twisted graphene bilayers. The possibility of tuning several twist angles in multilayer
graphene \cite{PhysRevX.9.031021,PhysRevResearch.2.033357,PhysRevB.100.085109,Carr2020,2020arXiv201213741W} suggests that these systems may realize correlated states of matter beyond the ones already observed in twisted bilayers.

\begin{figure}[!t]
    \centering
    \includegraphics[width=\columnwidth]{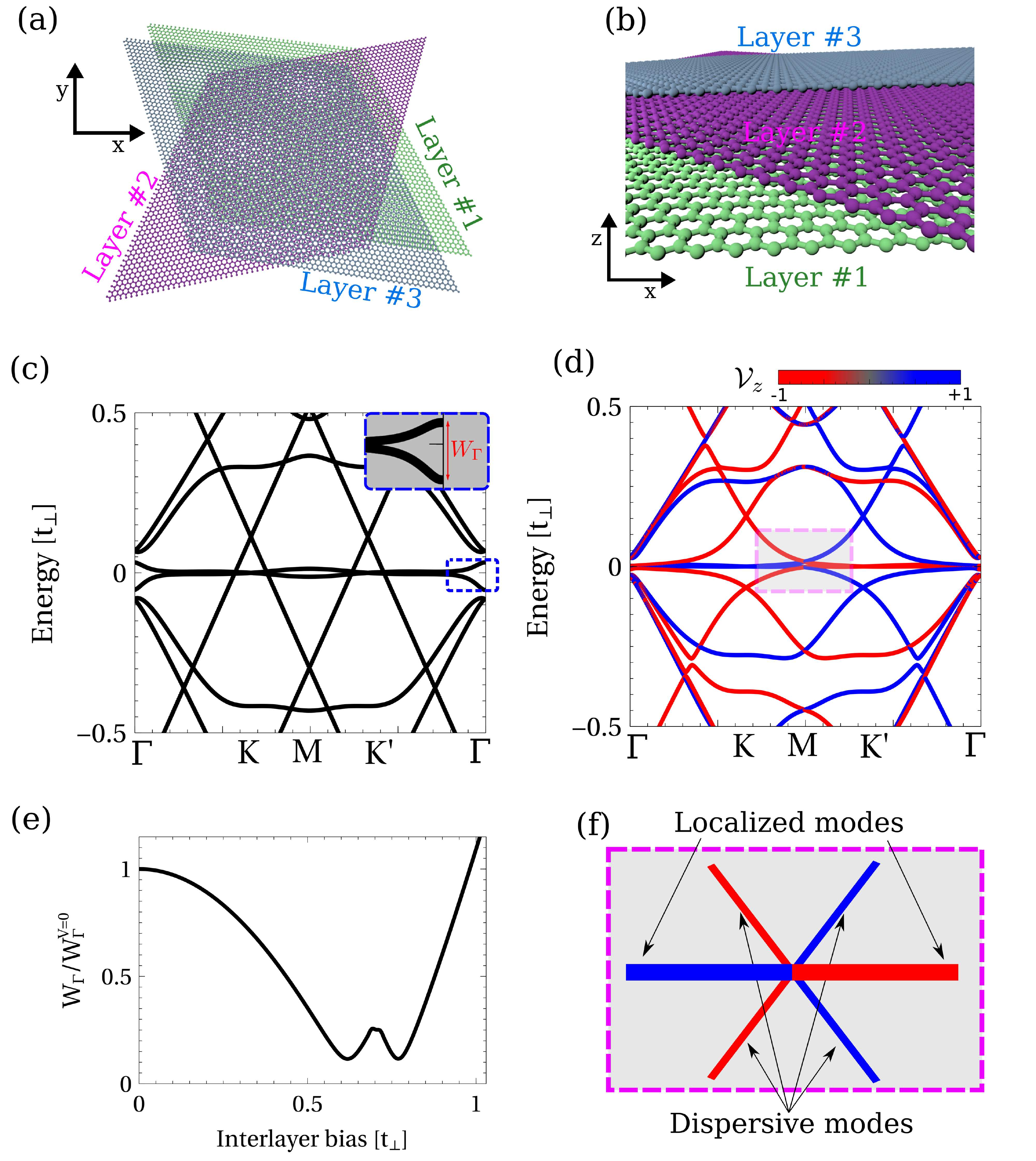}
        \caption{Top (a) and side (b) view of the twisted trilayer graphene system. Panel
        (c) and (d) display the electronic structure in the absence and presence of interlayer bias
        $V=0.6t_\perp$, respectively. The color code in (d) corresponds to the valley quantum number $\mathcal{V}_z$. Panel (e) shows how the dependence of the flat band width, defined in the inset of (c), on the interlayer bias. 
        Panel (f) schematically represents the low energy sector with localized and itinerant modes. 
	}
    \label{fig:fig1}
\end{figure}

Heavy fermions are a family of strongly correlated materials hosting a variety of fascinating quantum many-body states \cite{Coleman:2015,Wirth2016} 
and unusual quantum critical phenomena \cite{Si:2010,Ramires2019,SiGlobal,ColemanGlobal}. Two main ingredients constitute the electronic degrees of freedom in heavy fermions: localized states, usually associated with f-orbitals, and a sea of delocalized electrons. In these materials, strong interactions in the localized degrees of freedom generate local magnetic moments, which, through the Kondo effect, couple to the surrounding conduction electrons and dramatically renormalize their mass. Remarkably, graphene multilayers are known to host both dispersive \cite{PhysRevB.88.121408,Rickhaus2018} and localized electronic states \cite{PhysRevB.82.121407,correlatedCao2018,Xie2019}. 
However, the coexistence of dispersive and localized
modes does not occur in the most studied instances
of twisted graphene bilayers and bi-bilayers.

Here we establish that twisted graphene trilayers can host electronic states that mimic heavy fermion systems in a purely carbon-based material. We show that the main ingredients, flat bands, and highly dispersive states, appear simultaneously and provide the starting point to engineer and control electronic phases analogous to the ones observed in heavy fermions. We discuss how, in the presence of interactions, 
the highly degenerate flat bands lead to the formation of local spin and valley moments, which become exchange coupled to dispersive helical modes, leading to the emergence of heavy fermions. Our proposal allows to
electrically explore a Doniach  phase  diagram, putting forward a tunable platform
heavy-fermion physics in van der Waals materials.

We start by introducing the atomistic Hamiltonian for the twisted trilayer,  demonstrating that these systems host the fundamental ingredients to realize heavy fermion physics. We take a single orbital per carbon atom, yielding a tight-binding Hamiltonian of the form \cite{PhysRevB.82.121407,PhysRevB.92.075402}

\def\br{{\bold{r}}}

\begin{eqnarray}
\label{eq:htb}
        \mathcal{H} = 
	-t \sum_{\langle i,j \rangle,s} c_{i,s}^\dagger 	c_{j,s} -
        \sum_{i,j,s} 
	\bar t_{\perp} (\br_i,\br_j) c_{i,s}^\dagger c_{j,s} \\ \nonumber
		+ \frac{V}{2d} \sum_{i,s} z_i c_{i,s}^\dagger c_{i,s},
\end{eqnarray}
where $c_{i,s}^\dagger (c_{i,s})$ creates (annihilates) electrons at site $i$ with spin $s$. In the first term, $t$ is the intralayer hopping amplitude, and $\langle i,j \rangle$ restricts the sum to first neighbors. The second term accounts for interlayer hopping defined as
 $
\bar t_{\perp}(\bold r_i,\bold r_j) =
t_{\perp}
\frac{(z_i - z_j)^2 }{|\bold r_i - \bold r_j|^2}
e^{-\beta (|\bold r_i - \bold r_j|-d)}$ \cite{PhysRevB.82.121407,PhysRevB.92.075402}, 
where $d$ is the interlayer distance and $\beta$ controls the spatial decay of the hopping amplitudes. As a reference, for twisted graphene multilayers
$t \approx 3$ eV and $t_\perp \approx 0.15 t$ \cite{RevModPhys.81.109}. The third term corresponds to interlayer bias with magnitude $V$, with $z_i=\pm d$ for the upper/lower layer and $z_i=0$ for the middle layer. Here we consider a twisted trilayer structure with the top and bottom layers aligned, rotated by an angle $\theta$ with respect to the middle layer, as shown in Fig.~\ref{fig:fig1} (a) and (b).

The electronic structure for $\theta = 1.6^\circ$ is shown in Fig.~
\ref{fig:fig1} (c) for $V=0$. We highlight the emergence of flat bands,
similar to the ones observed in twisted bilayer graphene at the magic angle, coexisting with dispersive modes \cite{PhysRevResearch.2.033357,PhysRevB.100.085109,Carr2020,2020arXiv201213741W,2019arXiv190712338L}. 
Interestingly, when applying an interlayer bias, this twisted
system develops even flatter bands, as shown in Fig.~\ref{fig:fig1} (d), in stark contrast with the case of magic-angle bilayers, for which an interlayer bias does not impact the electronic structure strongly \cite{Yankowitz2019}.
This can be quantitatively
assessed by computing the splitting
of the nearly flat bands at the $\Gamma$ point as a function
of the interlayer bias,
as shown in Fig.~\ref{fig:fig1} (e). 
In particular, it is observed that for interlayer
biases of $V\approx 0.5 t_\perp$ the bandwidth becomes
drastically reduced. Moreover, in the presence of interlayer bias, the electronic states can still be associated with a well-defined valley quantum number $\mathcal{V}_z$ (computed in the
real-space basis using the valley operator \cite{PhysRevLett.120.086603,Ramires:2018,PhysRevB.99.245118,PhysRevLett.123.096802,PhysRevLett.126.056803}). In summary, we have the coexistence of valley polarized and spin degenerate ultra flat states and dispersive modes, as shown schematically in Fig.~\ref{fig:fig1} (f).

\begin{figure}[t!]
    \centering
    \includegraphics[width=\columnwidth]{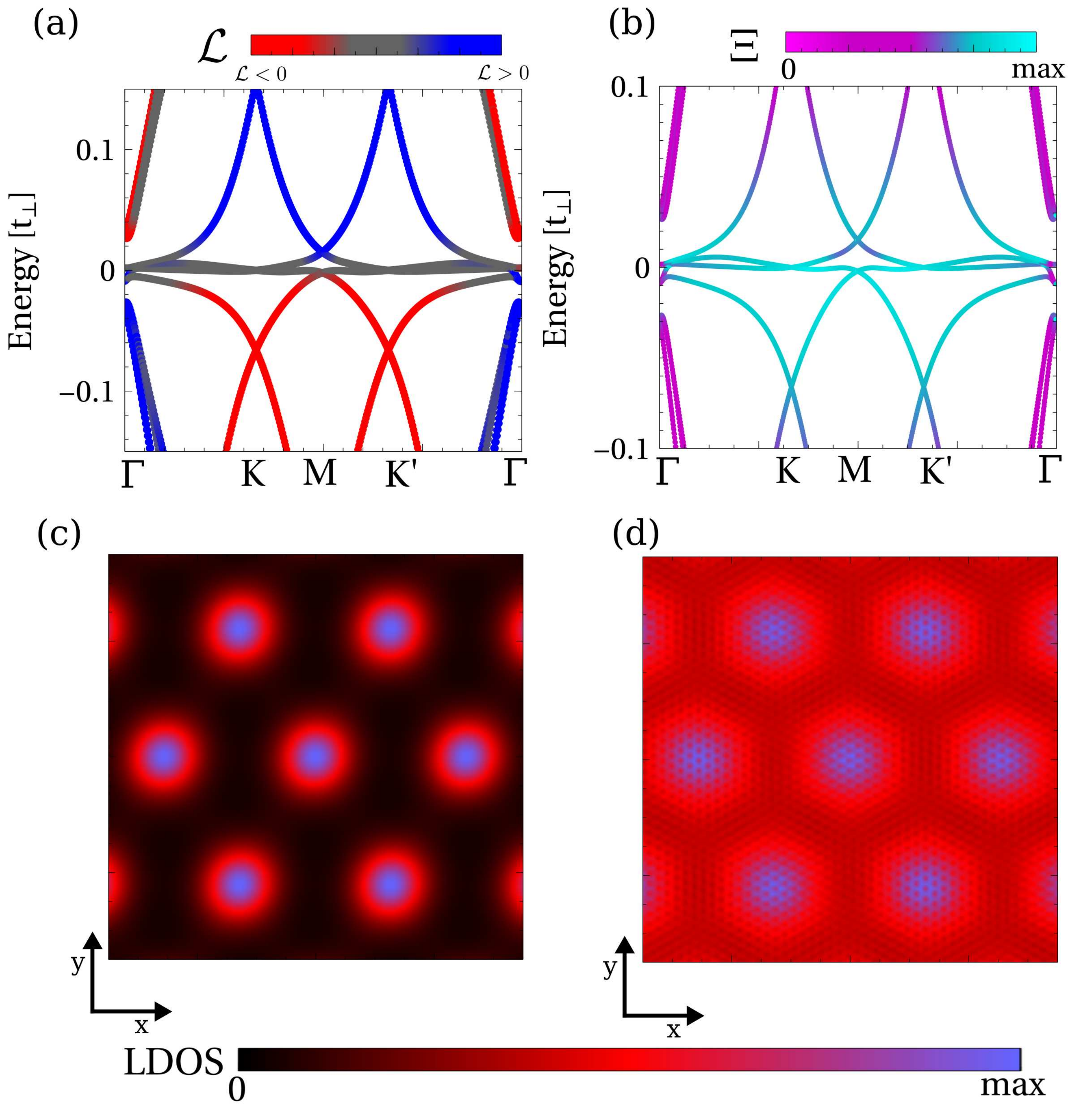}
        \caption{(a,b) Band structure
	of the biased ($V=0.6t_\perp$) twisted trilayer,
	highlighting the layer polarization
	of the states, $\mathcal{L}$, in (a), and
	their localization in the unit cell, quantified by the inverse participation ratio $\Xi$, in (b).
	Panels (c,d) show the local density of states
	 for the flat modes
	 at $\omega=0$ (c) and for the
	 dispersive modes at
	 $\omega=0.03t_\perp$ (d). 
        }
    \label{fig:fig2}
\end{figure}

\begin{figure}[t]
    \centering
    \includegraphics[width=\columnwidth]{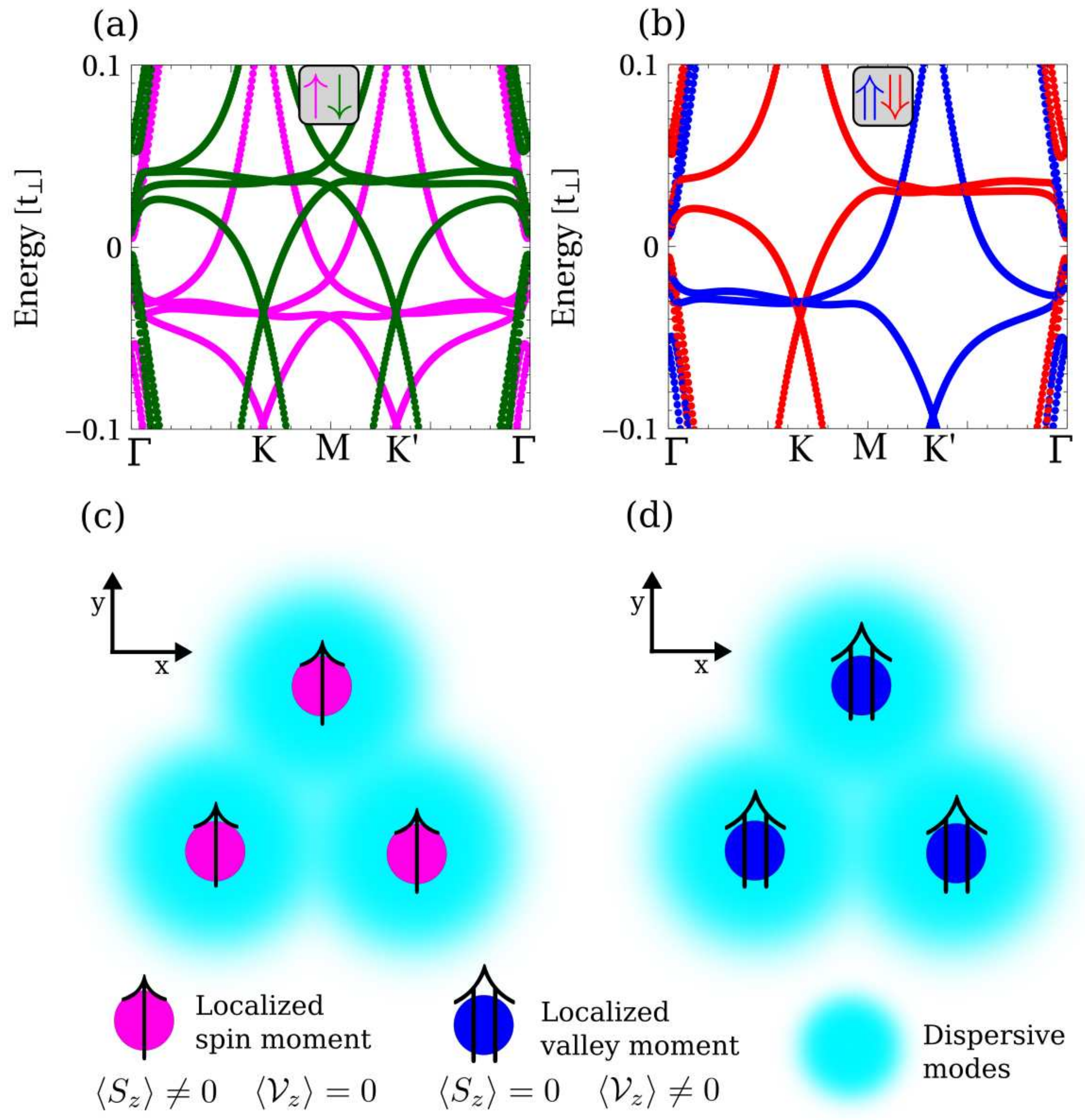}
        \caption{
		(a) Band-structure in the presence of finite
		spin (a) and valley (b) polarization of the
		localized modes.
		Panels (c) and (d) show a sketch of the order
		associated to the bandstructures in (a) and (b), respectively.
        }
    \label{fig:fig3}
\end{figure}

We now focus on the flattest regime in the presence of a bias, as shown in Fig.~\ref{fig:fig2}. 
By computing the layer polarization,  $\mathcal{L}$, shown in
Fig.~\ref{fig:fig2} (a), we conclude that  both the flat bands and dispersive modes are delocalized between the three layers, with dispersive modes showing a small layer polarization.
Moreover, the localization of the states in the moir\'{e} unit cell can be characterized by
means of the inverse participation ratio $\Xi = \sum_i |\Psi_n(i)|^4 $, Fig.~\ref{fig:fig2} (b). It is observed that the flat bands are strongly localized
in the moir\'{e} unit cell, whereas the dispersive modes become more delocalized. The different
localization can be directly observed from the local density of states (LDOS) at the energy of the
localized and dispersive modes, shown in Fig.~\ref{fig:fig2} (c) and (d), respectively.

To understand how the electronic structure
is modified in presence of interactions, we first study the problem at the mean-field level. This is equivalent to first-principles electronic structure
calculations
performed for heavy fermion compounds  \cite{PhysRevB.78.104113,Shick2001,PhysRevB.75.035109,PhysRevB.79.235125,PhysRevLett.60.2523,PhysRevLett.103.107202,PhysRevLett.118.197202}.
Electronic interactions for the atomistic Hamiltonian take the form
$\H_{\text{int}} = \sum_{ij} v(\mathbf r_i,\mathbf r_j) 
\rho_i \rho_j$, 
where $\rho_i = \sum_s c^\dagger_{i,s} c_{i,s}$,
and $v(\mathbf r_i,\mathbf r_j)$ is a screened Coulomb interaction.
A mean-field decoupling allows to transform the previous
density-density operator into a single-particle operator
of the form
$H^{MF} = \sum_{ijss'} \chi^{ss'}_{ij} c^\dagger_{i,s} c_{j,s'}$, where
$\chi^{ss'}_{ij}$ are the mean-field field parameters. 
In the following, we will consider two mean-field ansatzes, one leading
to a mean-field with spin polarization in the localized modes,
and one leading to valley polarization in the localized modes.
It must be noted that, although such mean-field electronic structure does not capture quantum fluctuations, it can be taken as an estimate of the backaction of the localized modes onto the dispersive states.
As a reference, in our calculations, we take an electronic
repulsion equivalent to a
charging energy of $50$ meV, comparable
to the one of twisted bilayer
graphene \cite{Guinea2018,PhysRevB.100.161102,PhysRevB.100.205113,Stepanov2020,2020arXiv200305050P}.
It is finally worth noting that
the strength of the Coulomb interactions can be externally controlled by means
of screening engineering through a substrate \cite{Stepanov2020,PhysRevB.100.161102}.

The mean-field band structures of the
atomistic model for the case of a finite spin and valley polarization are shown in Fig.~\ref{fig:fig3} (a) and (b), respectively, highlighting that the polarization of the
localized modes leads to strong and qualitatively distinct band reconstructions. This phenomenology is analogous to the one found in twisted bilayers \cite{Guinea2018,2020arXiv200305050P,PhysRevB.100.205113},
with the key difference that twisted bilayers do not have dispersive modes
coexisting at the Fermi energy.
In the proposed twisted trilayer graphene system, these results highlight that the dispersive modes
are strongly affected by the localized ones, signaling the existence of exchange coupling between extended and localized modes,
what can be understood in terms of a Kondo coupling \cite{Liechtenstein1987,Oswald1985,PhysRevB.82.180404,PhysRevLett.111.127204}.

The phenomenology obtained from the microscopic atomistic model within a mean-field approach shows that biased twisted trilayers
have the fundamental ingredients to display heavy fermion physics. 
In order to proceed with the discussion, 
we now focus on the
low energy effective model for  twisted trilayer graphene, captured by:

\begin{equation}
H= H_{Helical}+H_{LM}+H_{Hyb}+H_{Int}.
	\label{eq:h}
\end{equation}
Here $H_{Helical}$ corresponds to the propagating helical dispersive modes.
Note that, 
in contrast to the usual concept of helicity locking the direction of propagation with 
the spin projection, here the helical modes have the direction of propagation 
locked with the valley DOF, while they are spin-degenerate:
\begin{eqnarray}
H_{Helical} 
&\approx& v_F \sum_{k>0,\sigma} k\dis_{k \Uparrow \sigma}^\dagger \dis_{k \Uparrow \sigma} + \\ \nonumber
& & v_F \sum_{k<0,\sigma} k\dis_{k \Downarrow \sigma}^\dagger \dis_{k \Downarrow \sigma},
\end{eqnarray}
where $\sigma=\{\uparrow,\downarrow\}$ and $v=\{\Uparrow,\Downarrow\}$ correspond to the spin and valley
degrees of freedom (DOF), respectively,
$k$ denotes the wavevector difference with respect to the
Fermi surface, $v_F$ stands for 
the Fermi velocity. The flat bands are accounted as localized states at each moir\'{e} site with internal spin and valley DOFs:
\begin{eqnarray}
H_{LM} = \sum_{i,\sigma,v} \epsilon_{LM} f_{i v \sigma}^\dagger f_{i v \sigma}.
\end{eqnarray}
From the atomistic calculations, Fig. \ref{fig:fig1} (b), we access that the hybridization between the localized and helical modes preserves spin and valley DOF, therefore we write:
\begin{eqnarray}
H_{Hyb} = \delta \sum_{i,\sigma,v} f_{iv \sigma}^\dagger \dis_{i v \sigma}+h.c.,
\end{eqnarray}
where $\delta$ is the hybridization strength. As a reference, our calculations
show an effective hybridization
strength of $\delta \approx 5$ meV. The interaction Hamiltonian penalizes multiple occupancy of the flat band modes
\begin{eqnarray}
H_{Int} = U \sum_{i, v, v',\sigma, \sigma'} n_{if,v\sigma} n_{if,v'\sigma'},
\end{eqnarray}
where $U$ is a generalized Hubbard interaction parameter. The sum should exclude terms with both $v=v'$ and $\sigma=\sigma'$. 
Interactions in the dispersive modes would lead to a renormalization of the Fermi velocity,
and are reabsorbed in $H_{Helical}$.\cite{Gonzlez1994,RevModPhys.84.1067,Elias2011,PhysRevLett.98.216801,PhysRevLett.99.226803,PhysRevLett.118.266801}.
We note that it is possible
to write more generic 
inter-orbital coupling,
yet SU(4)
models are known to effectively
account for the interactions of localized modes
in graphene multilayers \cite{PhysRevLett.122.246401,PhysRevLett.122.246402}.

We can now find an effective Kondo model by integrating out the the high energy
DOF of the generalized Anderson model. This can be done through a
Schrieffer-Wolf transformation. 
In particular, from our atomistic calculations we obtain
$U=50$ meV and $\delta \approx 5$ meV, which fulfill the $\delta \ll U$ criteria
required to map the Anderson model to the Kondo model.
Note that if the assumption of a spin and
valley conserving hybridization holds, we can assign a unique index $\alpha =
(v, \sigma)$ summed over four flavours, such that the problem can be casted as
a SU(4) Kondo model
\cite{Borda:2003,Zarand:2003,Lopez:2005,Choi:2005,Filippone:2014,PhysRevB.102.104514}:
\begin{eqnarray}
H_{Kondo} &=& H_{Helical} + J_K\sum_{i} \bold{S}_i \cdot \bold{s}_i,
\end{eqnarray}
where $J_K\approx \delta^2/U$ 
is the Kondo coupling, $\bold{S}_i = \Psi_{fi}^\dagger \boldsymbol{\Gamma} \Psi_{fi}$, and $\bold{s}_i = \Psi_{di}^\dagger \boldsymbol{\gamma} \Psi_{di}$. Here $\boldsymbol{\Gamma}$ and $\boldsymbol{\gamma}$ stand for the array of 15
generators of SU(4) plus the identity, and
$\Psi_{fi}^\dagger=(f^\dagger_{i\Uparrow\uparrow},f^\dagger_{i\Uparrow\downarrow},f^\dagger_{i\Downarrow\uparrow},f^\dagger_{i\Downarrow\downarrow})$
and $\Psi_{di}^\dagger=(\dis^\dagger_{i\Uparrow\uparrow},\dis^\dagger_{i\Uparrow\downarrow},
\dis^\dagger_{i\Downarrow\uparrow},\dis^\dagger_{i\Downarrow\downarrow})$.
Interestingly, the effect of enlarging the symmetry group associated with the
degrees of freedom that can scatter through the Kondo coupling by generalizing
the Kondo effect from SU(2) to SU(N) enhances the Kondo temperature according to 
$
T_K^{SU(N)} \approx D (N\rho_0 J_K)^{1/N} e^{1/(N \rho_0 J_K)}
$, 
where $\rho_0$ is the flat density of states and $D$ the bandwidth associated with the helical modes \cite{Choi:2005,Filippone:2014}. Note that the Kondo temperature is exponentially enhanced by
increasing $N$ \cite{PhysRevB.28.5255,PhysRevB.93.035120,PhysRevLett.57.877,PhysRevB.35.3394,PhysRevB.38.316}.
The fact that the delocalized states 
are helical does not play a critical role in the
Kondo effect and on the renormalization procedure
as long as the DOS is finite \cite{Mitchell:2013,Zitko:2010}.

\begin{figure}[t]
    \centering
    \includegraphics[width=\columnwidth]{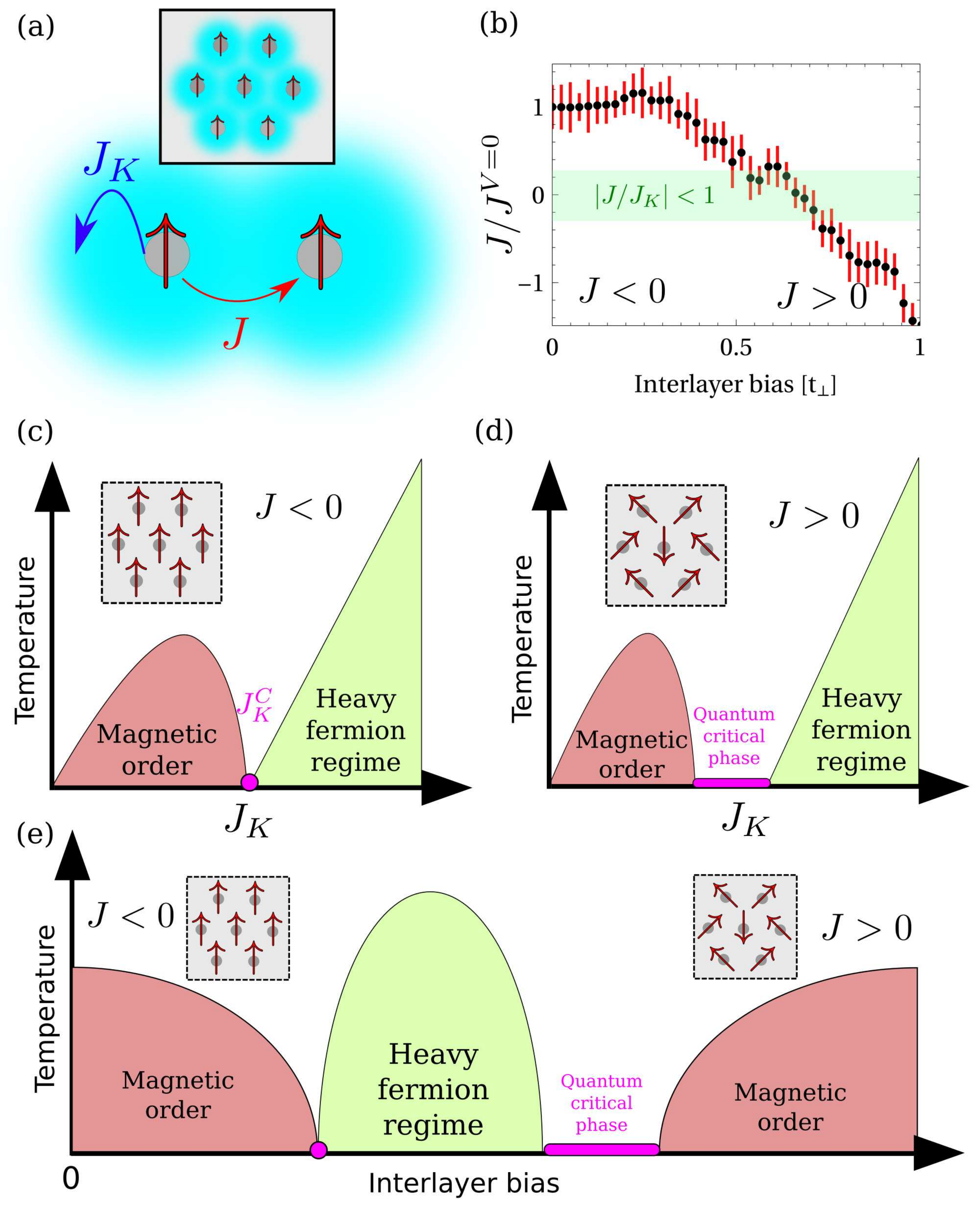}
        \caption{
		(a) Sketch of the exchange between local moments
		of strength $J$, and the Kondo coupling between
		extended and localized modes of strength $J_K$.
		Panel (b) shows the dependence of the exchange coupling
		with the interlayer bias, showing an electrical switch
		from ferromagnetic to antiferromagnetic.
		Green region in (b) shows
		the regime in which $|J/J_K|<1$.
		Panel (c) shows the phase diagram in the presence
		of ferromagnetic exchange and
		panel (d) in the presence of antiferromagnetic exchange.
		Panel (e) shows how the three phases can be explored by
		applying an interlayer bias between the layers, stemming
		from the exchange control shown in (b).
        }
    \label{fig:fig4}
\end{figure}

The existence of the heavy-fermion regime requires that
Kondo interaction, $J_K$, dominates the exchange coupling
between localized
moments $J$. As a reference, our atomistic
calculations yield
an effective Kondo coupling
of the order of $J_K \approx 0.5$ meV.
In the SU(4) scenario, the exchange coupling between
local modes takes the form
$H_{Exc} = J\sum_{\langle i j \rangle} \bold{S}_i
\cdot \bold{S}_{j}$.
Interestingly, exchange couplings in twisted
van der Waals materials 
have been shown  
to be tunable all the way from ferromagnetic
to antiferromagnetic
\cite{PhysRevLett.119.107201,PhysRevLett.126.056803,PhysRevX.8.031087}.
 We now compute the exchange $J$ between two neighbouring localized modes by means of the magnetic
force theorem \cite{Liechtenstein1987,Oswald1985,PhysRevB.82.180404,PhysRevLett.111.127204}. 
In absence of interlayer bias, we obtain
a ferromagnetic coupling between local moments of
$|J(V=0)| \approx 2$ meV \footnote{The effective exchange couplings
depends on the charging energy, that can be controlled via dielectric engineering}.
Once an interlayer bias is applied, the exchange coupling is reduced and can even change sign for 
increasing bias, as shown in Fig.~\ref{fig:fig4} (b). In particular, we obtain that the exchange between local
moments can be switched off for $V \approx 0.5 t_\perp$.
In this regime, the Kondo coupling
$J_K$ is the dominant interaction, driving the
twisted trilayer to the heavy-fermion regime.
Importantly, depending on the sign of $J$, the phase diagram
of the twisted trilayer will take two different forms.
For ferromagnetic coupling, $J<0$,
a quantum critical point separates the magnetically ordered phase
from the Kondo screened phase, Fig. \ref{fig:fig4} (c). In comparison, for
antiferromagnetic coupling, 
$J>0$, the geometric frustration of the
superlattice potentially leads to a quantum critical phase for a finite range of couplings \cite{Zhao2019,Ramires2019,PhysRevB.97.235117,Friedemann2009}, as shown in Fig. \ref{fig:fig4} (d).

Interestingly, the electric tunability of the twisted trilayer allows for the exploration of these two types of phase diagram with a single control parameter, the interlayer bias, as displayed in (Fig.~\ref{fig:fig4} (e)). This finding proposes a new perspective on the global phase diagram for heavy fermions \cite{SiGlobal,ColemanGlobal}.
While a precise first-principles estimate of all the couplings
is challenging  \cite{PhysRevResearch.2.033357},
the exact system's parameters can be inferred from experiments 
looking at the evolution of the tunneling
spectra as a function of temperature and magnetic
fields in different directions.
In particular, below the coherence temperature, 
Fano-like resonances appear and
the characteristic parameters of the line-shape can be extracted and identified with model parameters such as the hybridization between localized and delocalized states \cite{Aynajian2012,Maltseva:2009,Figgins:2010,Schmidt2010, Ernst2011}. 

To summarize, we have shown, combining accurate
atomistic calculations and low energy effective models, 
that twisted graphene trilayers provide a
van der Waals platform to engineer heavy fermion physics. In particular,
we demonstrated that the existence of electrically controllable flat bands
and dispersive modes provides the single-particle
starting point to simulate Kondo lattices. In the presence of
interactions, the flat bands lead to local spin and valley moments. 
Interestingly, the electric tunability of the inter-moment exchange couplings allows for the exploration of both the conventional
and frustrated Kondo lattice regimes, such that twisted trilayers can realize the global Doniach phase diagram for heavy fermions within a single material platform. 
Our results show that twisted graphene
multilayers provide a carbon-only platform to emulate rare earth compounds,
opening new possibilities
in the field of correlated twisted van der Waals materials.

\begin{acknowledgments}
\textit{Acknowledgments:}
A. R. acknowledges
the financial support from SNSF
Ambizione.
J. L. L. acknowledges
the computational resources provided by
the Aalto Science-IT project,
the
financial support from the
Academy of Finland Projects No.
331342 and No. 336243, and the
Jane and Aatos Erkko Foundation.
We thank M. Sigrist, T. Neupert, P. Liljeroth and
P. Rickhaus for useful discussions.
\end{acknowledgments}


\bibliography{biblio}{}

\end{document}